\documentclass[twocolumn,amssymb, nobibnotes, superscriptaddress, aps, prd]{revtex4}
\usepackage{amsmath}
\usepackage{epsfig}
\usepackage{graphicx}% Include figure files
\usepackage{subfigure}% Include figure files
\usepackage{dcolumn}% Align table columns on decimal point
\usepackage{bm}% bold math
\usepackage[bookmarks=true,
   colorlinks=true,
   linkcolor=blue,
   urlcolor=blue,
   citecolor=blue,
   bookmarks=true,
   hyperindex=true
]{hyperref}
\usepackage{graphicx} %use graph format
\usepackage{epstopdf}
\usepackage{siunitx}
\usepackage{bookmark}
\usepackage{amsmath}

\begin{document}

%Title of paper
\title{Restoring Velocity Immunity via Dynamic Mirror Compensation in a Large-Area Dual-Atom-Interferometer Gyroscope}

\author{Jie Gu}
\thanks{These authors contributed equally to this work}
\affiliation{Wuhan Institute of Physics and Mathematics, Innovation Academy for Precision Measurement Science and Technology, Chinese Academy of Sciences, Wuhan 430071, China}
\affiliation{University of Chinese Academy of Sciences, Beijing 100049, China}
\affiliation{Hefei National Laboratory, Hefei 230088, China}
\author{Yin-fei Mao}
\thanks{These authors contributed equally to this work}
\affiliation{Wuhan Institute of Physics and Mathematics, Innovation Academy for Precision Measurement Science and Technology, Chinese Academy of Sciences, Wuhan 430071, China}
\affiliation{University of Chinese Academy of Sciences, Beijing 100049, China}
\affiliation{Hefei National Laboratory, Hefei 230088, China}
\author{Zhan-Wei Yao}
\email[]{yaozhw@apm.ac.cn}
\affiliation{Wuhan Institute of Physics and Mathematics, Innovation Academy for Precision Measurement Science and Technology, Chinese Academy of Sciences, Wuhan 430071, China}
\affiliation{Hefei National Laboratory, Hefei 230088, China}
\author{An-qing Zhang}
\affiliation{Wuhan Institute of Physics and Mathematics, Innovation Academy for Precision Measurement Science and Technology, Chinese Academy of Sciences, Wuhan 430071, China}
\affiliation{University of Chinese Academy of Sciences, Beijing 100049, China}
\affiliation{Hefei National Laboratory, Hefei 230088, China}
\author{Si-Bin Lu}
\affiliation{Wuhan Institute of Physics and Mathematics, Innovation Academy for Precision Measurement Science and Technology, Chinese Academy of Sciences, Wuhan 430071, China}
\author{Shao-kang Li}
\affiliation{Wuhan Institute of Physics and Mathematics, Innovation Academy for Precision Measurement Science and Technology, Chinese Academy of Sciences, Wuhan 430071, China}
\author{Min Jiang}
\affiliation{Wuhan Institute of Physics and Mathematics, Innovation Academy for Precision Measurement Science and Technology, Chinese Academy of Sciences, Wuhan 430071, China}
\author{Xiao-Li Chen}
\affiliation{Wuhan Institute of Physics and Mathematics, Innovation Academy for Precision Measurement Science and Technology, Chinese Academy of Sciences, Wuhan 430071, China}
\author{Min Ke}
\affiliation{Wuhan Institute of Physics and Mathematics, Innovation Academy for Precision Measurement Science and Technology, Chinese Academy of Sciences, Wuhan 430071, China}
\author{Xi Chen}
\affiliation{Wuhan Institute of Physics and Mathematics, Innovation Academy for Precision Measurement Science and Technology, Chinese Academy of Sciences, Wuhan 430071, China}
\affiliation{Wuhan Institute of Quantum Technology, Wuhan 430206, China}
\author{Run-Bing Li}
\email[]{rbli@wipm.ac.cn}
\affiliation{Wuhan Institute of Physics and Mathematics, Innovation Academy for Precision Measurement Science and Technology, Chinese Academy of Sciences, Wuhan 430071, China}
\affiliation{Hefei National Laboratory, Hefei 230088, China}
\affiliation{Wuhan Institute of Quantum Technology, Wuhan 430206, China}
\author{Jin Wang}
\affiliation{Wuhan Institute of Physics and Mathematics, Innovation Academy for Precision Measurement Science and Technology, Chinese Academy of Sciences, Wuhan 430071, China}
\affiliation{Hefei National Laboratory, Hefei 230088, China}
\affiliation{Wuhan Institute of Quantum Technology, Wuhan 430206, China}
\author{Ming-Sheng Zhan}
\email[]{mszhan@apm.ac.cn}
\affiliation{Wuhan Institute of Physics and Mathematics, Innovation Academy for Precision Measurement Science and Technology, Chinese Academy of Sciences, Wuhan 430071, China}
\affiliation{Hefei National Laboratory, Hefei 230088, China}
\affiliation{Wuhan Institute of Quantum Technology, Wuhan 430206, China}

\date{\today}

\begin{abstract}
% insert abstract here
We propose and demonstrate a dynamical mirror compensation scheme to restore velocity immunity in a large-area dual-atom-interferometer gyroscope. In an ideal Mach-Zehnder configuration, the phase shift is inherently immune to atomic velocity, but this property is broken by the Earth's rotation via the Coriolis effect. We overcome this by actively rotating the Raman mirrors during the pulse sequence to cancel the time-dependent angular offset. The implementation relies on a decouplable calibration-compensation chain to remove rotation-induced time-dependent terms. The scheme is validated on a dual-atom-interferometer gyroscope with an interference area of 21.1 cm$^2$. After compensation, the phase's dependence on atomic velocity is reduced 40-fold, and the velocity contribution to scale-factor stability is evaluated to be 0.13 ppm. The sensor achieves a rotation sensitivity of $1.3\times10^{-8}$ rad/s/Hz$^{1/2}$ and a stability of $1.9\times10^{-10}$ rad/s at 4500 s integration, together with a common-mode noise rejection ratio of up to 459, demonstrated in a seismic event. This work removes a key obstacle to scale-factor stabilization in atom-interferometer gyroscopes and paves the way for their applications in inertial navigation and geophysics.
\end{abstract}

\maketitle

High-precision gyroscopes are essential for fundamental physics, geodesy, and inertial navigation $\cite{Schreiber2013,Lefèvre2013}$. Among them, cold-atom-interferometer gyroscopes (CAIGs) are particularly promising because their scale factors can greatly exceed those of optical sensors, leading to advantages in sensitivity, accuracy, and long-term stability $\cite{Scully1993}$. To date, atomic gyroscopes have already demonstrated excellent sensitivities and long-term stabilities close to the state-of-the-art optical gyroscopes $\cite{Durfee2006a,Savoie2018}$.

Beyond sensitivity, scale-factor stability is critical for absolute rotation measurements at a fixed nonzero rate. For rotations on the order of Earth's rate, a scale-factor stability of 10$^{-9}$ is required for tests such as the Lense-Thirring effect $\cite{Hurst2017}$. The Sagnac phase is proportional to the scale factor, which is determined by the interference area $\cite{Anderson1994}$. In a Mach-Zenhder CAIG, whether of the light-pulse type (fixed interrogation time T) \cite{Gauguet2009a, Berg2015a, Yao2021} or the atom-beam type (fixed Raman-beam separation L) \cite{Gustavson2000, Kwolek2022, Meng2024}, the area is proportional to the atomic longitudinal velocity, and the phase consequently depends explicitly on velocity. However, for a free-falling atom in an inertial frame the velocity dependence of two interference paths cancels exactly, making the interference phase immune to atomic velocity. We refer to this property as velocity immunity.

In a laboratory fixed on the Earth's surface, Earth's rotation breaks this immunity. The Raman mirrors rotate with the Earth during the interferometer sequence, so the mirror-atom relative geometry is different at each pulse. This opens the interference loop in the atom's view and introduces additional phase terms coupling to the atomic position and velocity $\cite{Lan2012,Asenbaum2020}$. Even for laser-cooled clouds, whose velocity instability is at the ppm level $\cite{Gauguet2009a,Chen2024}$, these extra terms can cause scale-factor fluctuations of several ppm, which is comparable to the tidal variations that affect a four-pulse configuration \cite{Stockton2011, Gautier2022, Xu2024}. Therefore, preserving velocity immunity under rotation requires a scheme that actively follows the Earth's motion.

\begin{figure*}[htbp]
	\centering
	\includegraphics[width=1\textwidth]{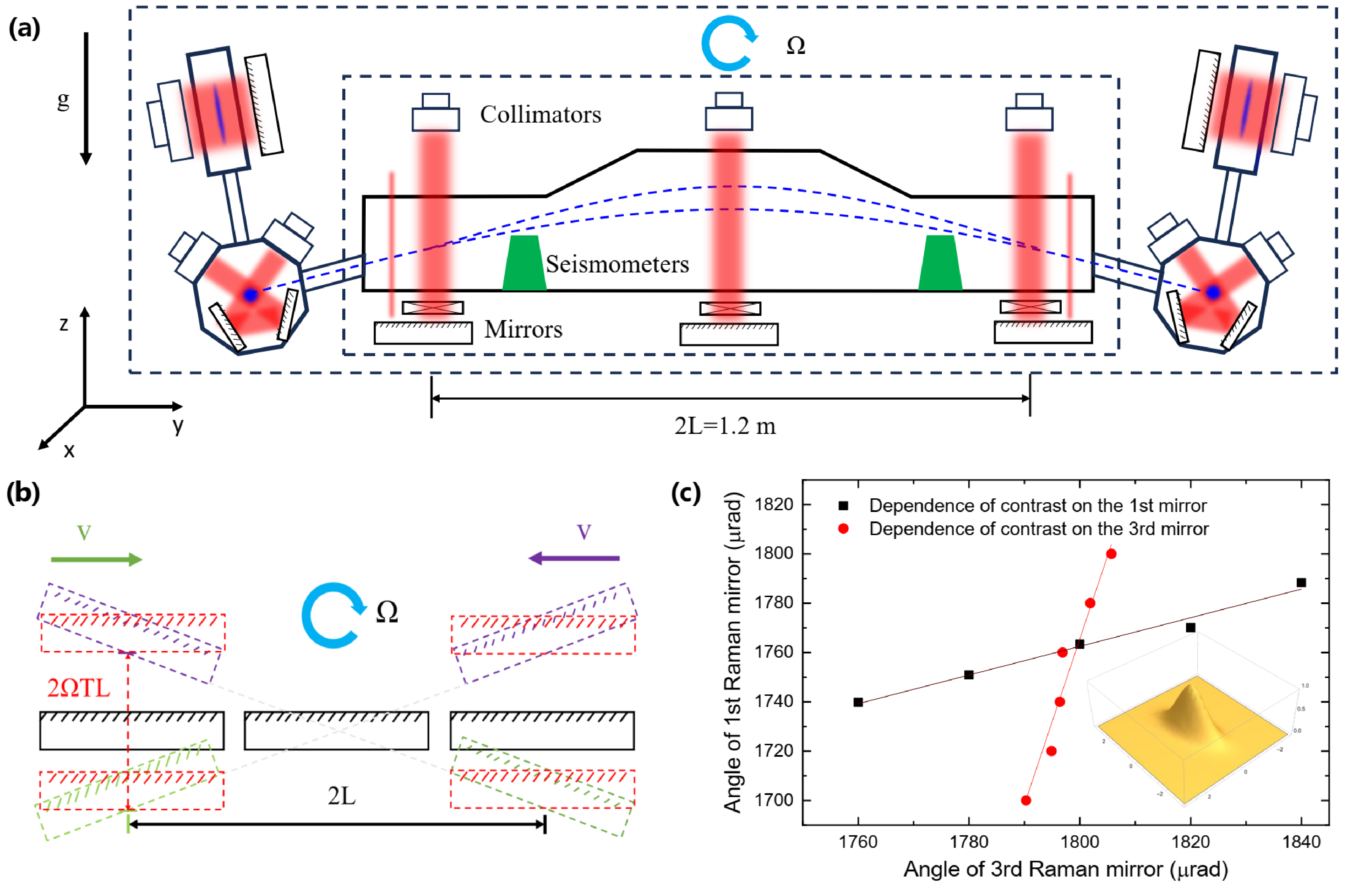}
	\caption{Principle of experimental. (a) Schematic of the dual large-area atom-interferometer gyroscope. (b) Dynamic mirror compensation scheme: black mirrors-Earth-fixed frame; green/purple-mirror angles seen by the 1st loop and 2nd one before compensation, and red-mirror angles after compensation in inertial frame. (c) Maximum contrast of the first (black squares) and third (red dots) mirror angle for a given setting of the other mirror, and the inset is the numerical simulation result.}\label{fig1}
\end{figure*}

In this Letter, we demonstrate that velocity immunity can be restored through a dynamic mirror compensation technique. By tilting the Raman mirrors during the interference pulse sequence, the atom can experience a constant effective mirror geometry, effectively rendering the system equivalent to being in an inertial frame. First, we identified that Earth's rotation reintroduces velocity coupling via a time-dependent mirror-atom relative angle, and that dynamic compensation is the necessary solution to restore immunity. Second, we developed a decouplable calibration-compensation chain: a contrast method for closing the interference loop, a phase method for calibrating  position-velocity coupling angles, and dynamic compensation for canceling rotation-induced time-dependent terms. Finally, in a dual-atom-interferometer gyroscope with an interference area of 21.1 cm$^2$, the velocity dependence is reduced by a factor of 40, the velocity contribution to scale-factor stability drops to 0.13 ppm, the sensitivity is improved to $1.3\times10^{-8}$ rad/s/Hz$^{1/2}$, and the stability achieves 1.9$\times10^{-10}$ rad/s at 4500 s.

The apparatus is shown in Fig. \ref{fig1}(a). $^{87}$Rb atoms are collected in a 2D magneto-optical trap (2D-MOT), transferred into a 3D MOT, and cooled to 4 $\mu$K by polarization-gradient cooling. They are launched at an angle of 65$^{\circ}$ relative to gravity using moving optical molasses, propagated at total velocity of 4.4 m/s, and then prepared in $\left| {F = 2,{m_F} = 0} \right\rangle $ via microwave selection and a blow-away laser. The Mach-Zehnder atom interferometer is realized with three counter-propagating Raman beam pairs ($\pi/2-\pi-\pi/2$) that couple $\left| {F = 1,{m_F} = 0} \right\rangle $ and $\left| {F = 2,{m_F} = 0} \right\rangle $. After interference the population in $\left| {F = 2,{m_F} = 0} \right\rangle $ is detected by fluorescence. To reject common-mode noise, two interferometers with nearly identical parabolic trajectories are operated simultaneously along opposite directions. Compared to our previous work $\cite{Yao2021}$, we have (i) extended the interrogation time to T=149.4 ms and the interference length to 0.6 m, giving an interference area of 21.1 cm$^2$---the largest reported for atom interferometers; (ii) integrated the most alignment-sensitive mirrors into the vacuum chamber to improve temperature stability; (iii) narrowed the detection laser linewidth to a few tens of kHz $\cite{Jiang2025}$, reducing detection noise to 15 mrad/shot for rotation phase, corresponding to a rotation noise of 7.3 nrad/s/Hz$^{1/2}$; (iv) implemented real-time vibration-noise compensation $\cite{Lu2025}$; and (v) relocated the gyroscope to a quieter cave laboratory, reducing the half-differential (corresponding to rotation signal) phase noise to less than 30 mrad per shot.

\begin{figure} [htbp]
	\flushleft
	\includegraphics[width=0.48\textwidth]{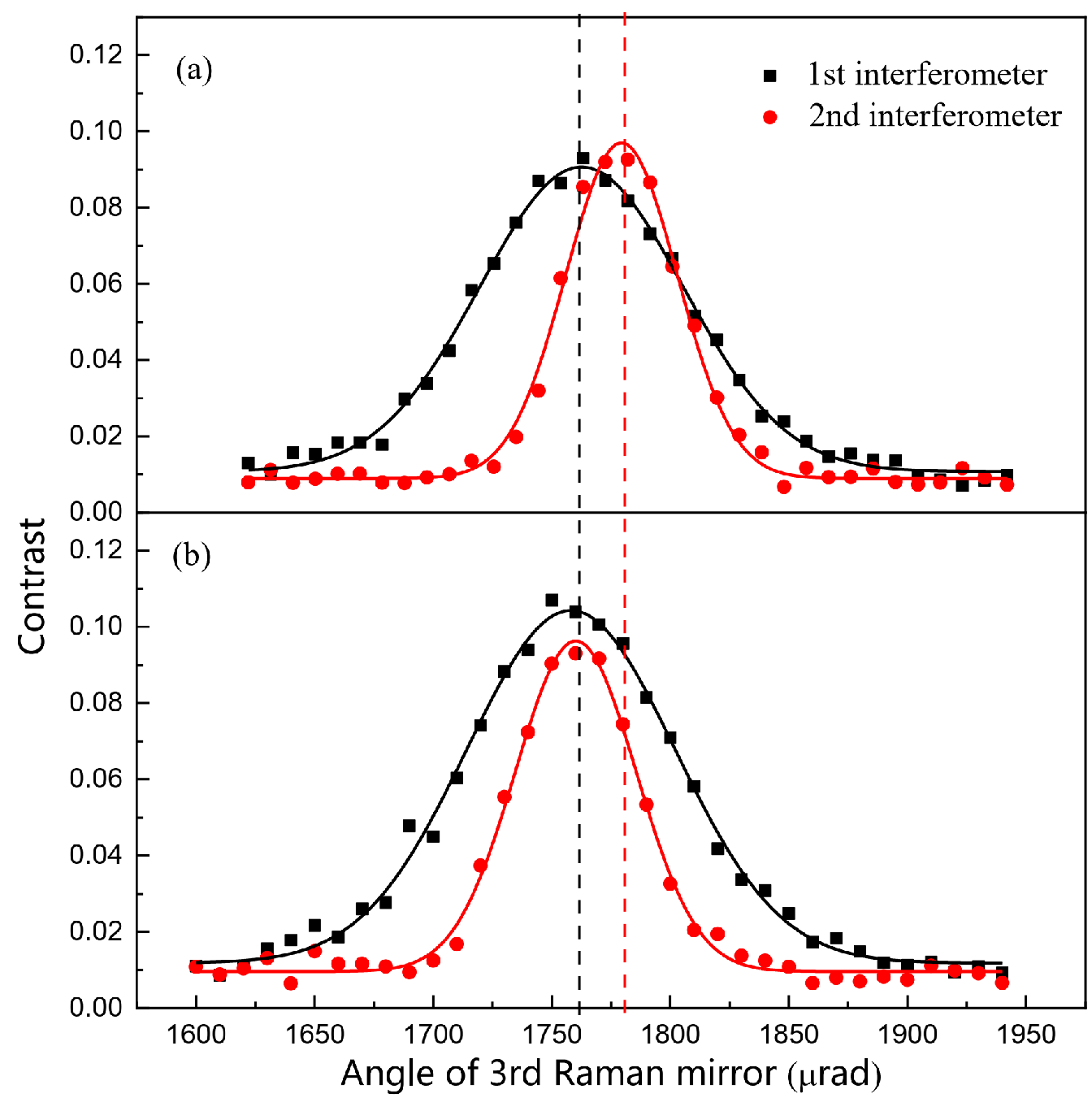}
	\caption{(color online) Fringe contrast versus mirror angle (a) before and (b) after dynamic compensation for the 1st loop (black squares) and 2nd one (red dots).}%
	\label{fig2}
\end{figure}

To realize velocity immunity in the large-area dual-interferometer system, we have developed a decouplable calibration-compensation chain.

$\textit{Stage 1: Contrast method---}$closing the interference loop. For interference to occur, the atomic trajectories must overlap within the coherence length $\cite{Lan2012,Roura2014}$ and no spatial phase gradients may be present $\cite{Tackmann2012,LiJ2025}$. Both conditions require mirror angles at the microradian level. The fringe contrast can be modeled as a two-dimensional Gaussian function of the angle deviations $\theta_1$ and $\theta_3$ (where $\theta_i$ is the relative angle between the ith mirror and the 2nd mirror):

\begin{equation}
	\begin{array}{l}
		C=e^{-(2\theta_1Tv_{\text{r}}/L_{\text{coh}} )^2}e^{-\frac{1}{2}[k_{\text{eff}}\sigma_{\text{y}}(\theta_1+\theta_3)]^2}\\
		 e^{-\frac{1}{2}[k_{\text{eff}}\sigma_{v_{\text{y}}}(\theta_1 t_{1}+\theta_3(t_1+2T))]^2},
	\end{array}
	\label{eq1}
\end{equation}
where $k_{\text{eff}}$ is the two-photon momentum transfer, $v_{\text{r}}$ is the recoil velocity, and $t_1$=51.2 ms is the time from launch to the first Raman pulse. By scanning $\theta_1$ and $\theta_3$ independently  and fitting the contrast maxima, as shown in Fig. \ref{fig1}(c), we extract the coherence length $L_{coh}$=109 nm, the initial cloud size $\sigma_{y}$=0.1 mm, and the velocity spread $\sigma_{v_y}$=5.0 mm/s. Thus, the proper angles for closing the interference loop correspond to the cross-point, as predicted by the numerical simulation results shown in the inset of Fig. \ref{fig1}(c). After alignment using the contrast method, the angles are determined at a few microradian level.

\begin{figure} [htbp]
	\flushleft
	\includegraphics[width=0.48\textwidth]{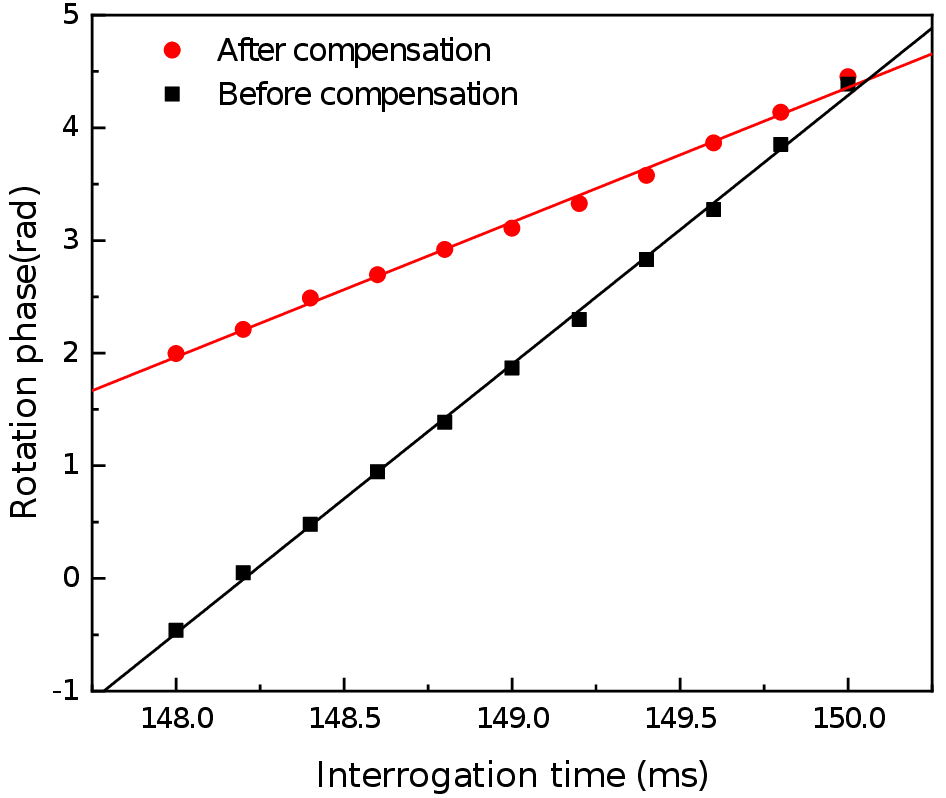}
	\caption{(color online) Rotation phase versus interrogation time T before (black squares) and after (red dots) dynamic compensation.}%
	\label{fig3}
\end{figure}

$\emph{Stage 2: Phase method---}$calibrating position-velocity coupling angles in interference loops. After coarse alignment, residual angle errors still couple to the atom's initial position and velocity, introducing a phase shift dependent on atomic position and velocity. For dual-loop atom interferometers with separated Raman beams, the corresponding phase shifts are expressed as

\begin{equation}
	\begin{array}{l}
		\delta {\varphi _{\rm{a}}}={k_{\rm{eff}}}[2{\Omega}LT+(y_{a}+v^{\prime}_{ya}t_{1})({\theta_{1a}}+{\theta_{3a}})+{2v^{\prime}_{ya}{\theta_{3a}}T]},
	\end{array}
\label{eq2}
\end{equation}
\begin{equation}
\begin{array}{l}
	\delta {\varphi _{\rm{b}}}={k_{\rm{eff}}}[-2{\Omega}LT+(y_{b}-v^{\prime}_{yb}t_{1})({\theta_{1b}}+{\theta_{3b}})-{2v^{\prime}_{yb}{\theta_{1b}}T}],
	\end{array}
\label{eq3}
\end{equation}
where y$_{a}$(y$_{b}$) and $v^{\prime}_{ya}(v^{\prime}_{yb})$ are the position and relative velocity (v$_{ya/b}$-L/T) for the 1st loop (2nd loop),  and $\theta_{1a}$, $\theta_{3a}$($\theta_{1b}$,  $\theta_{3b}$) are the mirror angles for the 1st loop (2nd loop) in the inertial frame, and they satisfy $\theta_{1a}$+2$\Omega$T=$\theta_{3b}$, $\theta_{1b}$+2$\Omega$T=$\theta_{3a}$ due to Earth's rotation. By modulating the first-pulse delay $t_{1}$ and the atomic velocity, we determine and null the coupling terms for each single loop, calibrating the position-velocity coupling angle to the sub-microradian level. Even with perfect static calibration in a single loop, the time-dependent angular offset 2$\Omega$T persists between the two loops as shown in Fig. \ref{fig1}(b) (green mirrors for the 1st loop and purple mirrors for the 2nd one), which make it impossible to decouple the velocity term simultaneously for dual loops. The contrast maxima for two interferometers cannot be reached simultaneously, as shown in Fig. \ref{fig2}(a). The angular deviation is 17.8 $\mu$rad, matching the Coriolis rotation angle 18.5 $\mu$rad expected from 2$\Omega_y$T$  \cite{Duan2020,Li2025}$.

$\textit{Stage 3: Dynamic compensation---}$restoring velocity immunity in dual interference loops. We eliminate the asymmetry by actively rotating the 1st and 3rd mirrors before each $\pi$/2 pulse, as shown in Fig. \ref{fig1}(b). Before the first $\pi$/2 pulse, the mirrors are set to the optimal angles for the 1st loop (up red mirrors) and 2nd loop (down red mirrors), respectively. Immediately after that pulse, we tilt both mirrors by 2$\Omega_y$T ($\approx$18.5 $\mu$rad) so that they adopt the optimal angles for each other. After dynamic compensation, both loops achieve maximum contrast simultaneously as shown in Fig. \ref{fig2}(b), confirming that the time-dependent Coriolis asymmetry has been removed. Under this condition, the rotation phase in Eqs. \eqref{eq2} and \eqref{eq3} simplifies to

\begin{equation}
	\begin{array}{l}
		{\varphi _{\Omega}} =2{k_{\rm{eff}}}{\Omega}LT,
	\end{array}
	\label{eq4}
\end{equation}
which no longer contains any atomic position or velocity terms. To verify the restoration of velocity immunity, we measure the rotation phase while varying the interrogation time T as shown in Fig. \ref{fig3}. Before compensation, the slope is 2.39$\pm$0.03 rad/ms (black squares), reflecting both rotation and spurious velocity couplings. After compensation, the slope decreases to 1.19$\pm$0.03 rad/ms (black dots), consistent with a purely linear $\phi_T\propto$T scaling as predicted by Eq. \eqref{eq4}. From the change in slope, we infer a reduction of the velocity dependence by a factor of 40, restoring the velocity immunity.

\begin{figure} [htbp]
	\includegraphics[width=0.48\textwidth]{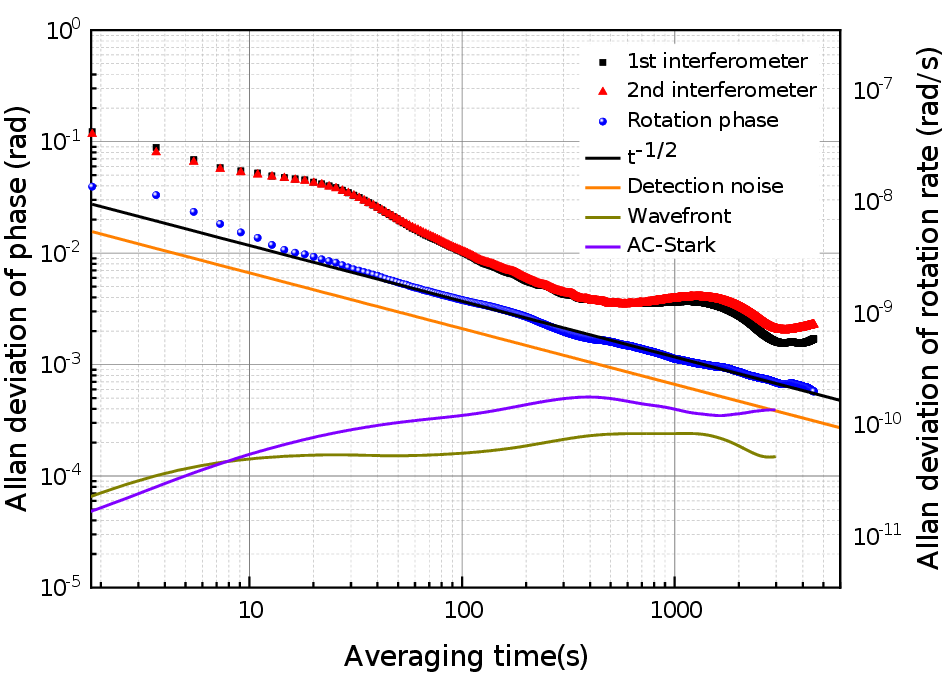}
	\caption{(color online) Allan deviations of the dual atom interferometers. Black squares: 1st interferometer; red triangles: 2nd interferometer; blue dots: rotation signal; black line: $\tau^{-1/2}$ scaling of 13 nrad/s/Hz$^{1/2}$; orange line: detection noise limit corresponding to 7.3 nrad/s/Hz$^{1/2}$; violet curve: phase drift caused by AC-Stark shifts; dark yellow curve: phase drift induced by wavefront distortions.}
\label{fig4}
\end{figure}

$\textit{Gyroscope performance.---}$The CAIG was evaluated by using Allan deviations, as shown in Fig. \ref{fig4}. The gyroscope operates with a cycle time of 0.91 s and achieves a sensitivity of 1.3$\times10^{-8}$ rad/s/Hz$^{1/2}$. The Allan deviation shows a stability of 1.9$\times10^{-10}$ rad/s after 4500 s. This long-term stability is a direct consequence of the improved scale factor: before compensation the same measurement gave a stability of 3$\times10^{-10}$ rad/s at a rotation rate of 12.4 $^{\circ}$/h, corresponding to a scale-factor instability of 5 ppm; after compensation, the velocity contribution to the scale-factor instability drops to 0.13 ppm. The remaining long-term instabilities are dominated by AC-Stark shifts ($\approx$0.4 mrad, violet curve) and wavefront distortions ($\approx$0.2 mrad, dark yellow curve) in Fig. \ref{fig4} $\cite{Altorio2020}$. These can be further reduced by stabilizing the Raman power $\cite{Lyu2022}$ and by using angle-sensitive laser interferometry to lock the mirror angles $\cite{Schuldt2009}$.

\begin{figure} [htbp]
	\flushleft
	\includegraphics[width=0.48\textwidth]{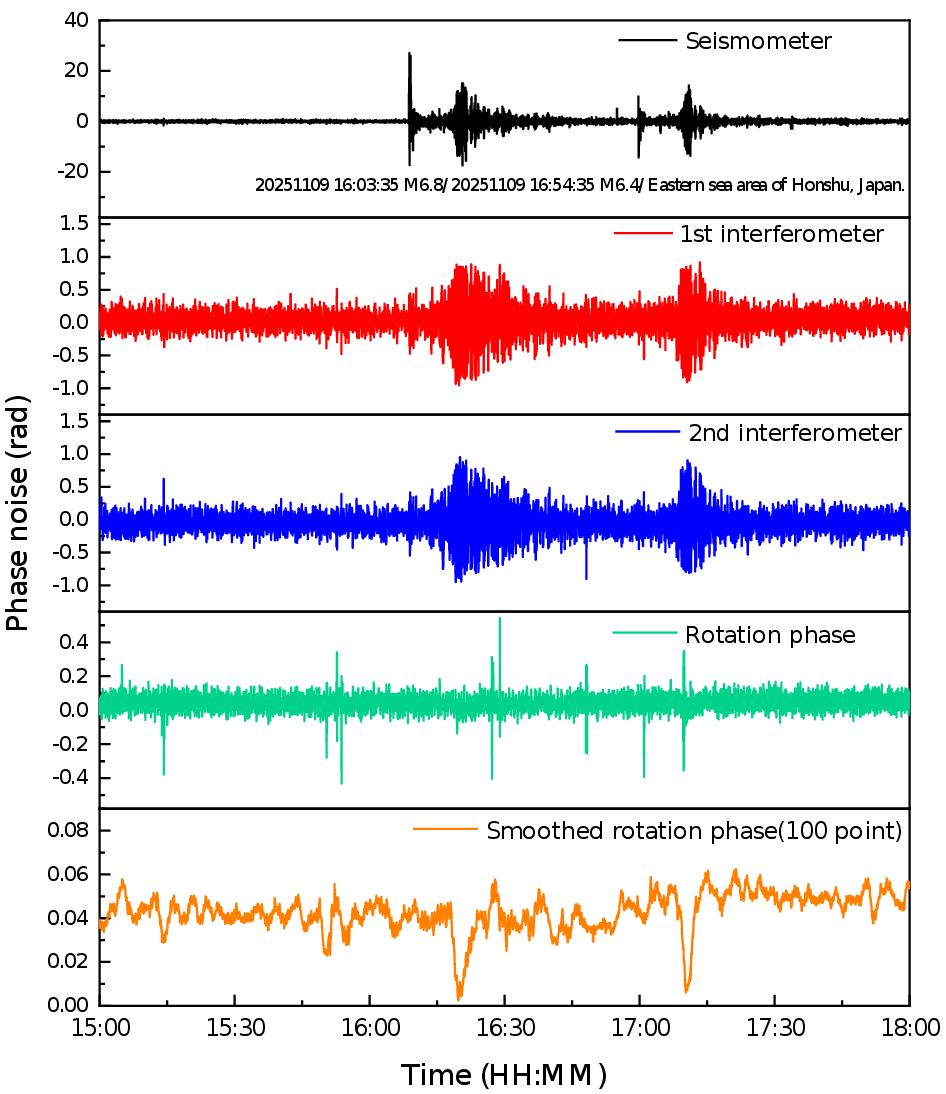}
	\caption{(color online) Seismic signal recorded by the dual atom interferometers. Black: seismometer weighted by the transfer function; red, blue: individual interferometer phases; green: rotation phase; orange: smoothed rotation phase.}%
	
\label{fig5}
\end{figure}

The CAIG was validated in seismic events. The dual-interferometer design provides strong common-mode rejection. Fig. \ref{fig5} shows the response to two earthquakes (magnitudes 6.8 and 6.4) that occurred in the east of Honshu, Japan, on 9 November 2025. Even though the seismic motion temporarily increases the deviation between the Raman mirrors and the reference seismometer, the two interferometer signals remain highly correlated. From the raw rotation data, we extract a common-mode rejection ratio of 459, proving the excellent symmetry of the system. In future, considering a duty cycle of 33$\%$, an interleaved operation  $\cite{Biedermann2013,Dutta2016,Savoie2018}$ can further improve the sensitivity to the nrad/s/Hz$^{1/2}$ level.  Furthermore, approaching the quantum projection noise limit would eventually enable rotation sensitivities at the sub-nrad/s/Hz$^{1/2}$ level $\cite{Gauguet2009a,Biedermann2009,Janvier2022}$, which means the seismic events could be detected more clearly and frequently.

$\textit{Conclusion.---}$We have demonstrated a dynamic mirror compensation method that restores the intrinsic velocity immunity of a Mach-Zehnder atom interferometer in the presence of Earth's rotation. The three-stage close-calibration-compensation chain provides a complete solution from coarse alignment of Raman mirrors to real-time cancellation of Coriolis effect. On a dual-atom-interferometer gyroscope with a record 21.1 cm$^2$ interference area, this approach reduces the velocity dependence by a factor of 40 and brings the velocity contribution to scale-factor stability to 0.13 ppm. The achieved sensitivity and long-term stability highlight the potential of this technique for precision gyroscopy, from geophysical monitoring to tests of fundamental physics.

We acknowledge the financial support provided by the  Quantum Science and Technology-National Science and Technology Major Project of China under Grant No. 2021ZD0300604; the National Key Research and Development Program of China under Grant No. 2025YFF0515200; the National Natural Science Foundation of China under Grants No. U25D9005 and No. 12504571; the Natural Science Foundation of Wuhan under Grant No. 2025040601030117. We also thank National Gravitation Laboratory in HUST for providing the experimental site for this study.

\bibliographystyle{plain}

\end{document}